# Automated screening of sickle cells using a smartphone-based microscope and deep learning

Kevin de Haan[†,1,2,3], Hatice Ceylan Koydemir[†,1,2,3], Yair Rivenson[1,2,3,*], Derek Tseng[1,2,3], Elizabeth Van Dyne[4], Lissette Bakic[5], Doruk Karinca[6], Kyle Liang[6], Megha Ilango[6], Esin Gumustekin[7], Aydogan Ozcan[1,2,3,8,*]

[1]Electrical and Computer Engineering Department, University of California, Los Angeles, CA, 90095, USA.

[2]Bioengineering Department, University of California, Los Angeles, CA, 90095, USA.

[3]California NanoSystems Institute (CNSI), University of California, Los Angeles, CA, 90095, USA.

[4] Department of Pediatrics, Division of Hematology-Oncology, David Geffen School of Medicine, University of California, Los Angeles, CA, 90095, USA.

[5]Department of Pathology and Laboratory Medicine, David Geffen School of Medicine, University of California, Los Angeles, Los Angeles, CA, USA

[6]Department of Computer Science, University of California, Los Angeles, CA, 90095, USA.

[7]Department of Neuroscience, University of California, Los Angeles, CA, 90095, USA.

[8]Department of Surgery, David Geffen School of Medicine, University of California, Los Angeles, CA, 90095, USA.

† Equally contributing authors

* Corresponding authors: rivensonyair@ucla.edu ; ozcan@ucla.edu

**Abstract**: Sickle cell disease (SCD) is a major public health priority throughout much of the world, affecting millions of people. In many regions, particularly those in resource-limited settings, SCD is not consistently diagnosed. In Africa, where the majority of SCD patients reside, more than 50% of the 0.2-0.3 million children born with SCD each year will die from it; many of these deaths are in fact preventable with correct diagnosis and treatment. Here we present a deep learning framework which can perform automatic screening of sickle cells in blood smears using a smartphone microscope. This framework uses two distinct, complementary deep neural networks. The first neural network enhances and standardizes the blood smear images captured by the smartphone microscope, spatially and spectrally matching the image quality of a laboratory-grade benchtop microscope. The second network acts on the output of the first image enhancement neural network and is used to perform the semantic segmentation between healthy and sickle cells within a blood smear. These segmented images are then used to rapidly determine the SCD diagnosis per patient. We blindly tested this mobile sickle cell detection method using blood smears from 96 unique patients (including 32 SCD patients) that were imaged by our smartphone microscope, and achieved ~98% accuracy, with an area-under-the-curve (AUC) of 0.998. With its high accuracy, this mobile and cost-effective method has the potential to be used as a screening tool for SCD and other blood cell disorders in resource-limited settings.

**Introduction**

Sickle cell disease (SCD) is the most common hematologic inherited disorder worldwide and a public health priority[1]. The majority of the world's SCD burden is in Sub-Saharan Africa, affecting millions of people at all ages. It is estimated that 200,000 to 300,000 children are born with SCD every year in Africa alone[2,3]. The prevalence of the disease varies across countries, being approximately 20% in Cameroon, Ghana and Nigeria and even rising up to 45% in some parts of Uganda[3].

SCD is an inherited disorder caused by a point mutation in hemoglobin formation, which causes the polymerization of hemoglobin and distortion of red blood cells in the deoxygenated state. As a result of this, the normally biconcave disc-shaped red blood cells become crescent or sickle shaped in people living with SCD. These red blood cells are markedly less deformable, have one-tenth the life span of a healthy cell, and can form occlusions in blood vessels. Children with SCD also suffer from spleen auto-infarction and the burden of disease becomes significant. Loosing splenic function, these children are at high risk for infections at an extremely young age, which significantly increases mortality rates[4]. Due to the lack of diagnosis and treatment, over 50% of these of children with SCD in middle and low-income countries will die[5].

Various methods have been developed for screening and diagnosis of SCD, including e.g., laboratory-based methods such as high performance liquid chromatography (HPLC)[6], isoelectric focusing[7], and hemoglobin extraction[8]. In addition to these relatively costly laboratory-based methods, there have been SCD diagnostic tests developed for point-of-care (POC) use[9–14]. These POC tests are mainly based on human reading, and human errors along with the storage requirements of these tests (involving e.g., controlled temperature and moisture to preserve chemical activity/function) partially limit their effectiveness to screen SCD, especially in resource limited settings[14].

An alternative method used for screening of SCD involves microscopic inspection of blood smear samples by trained personnel. In fact, each year hundreds of thousands of blood smear slides are prepared in sub-Saharan Africa to make diagnosis of blood cell infections and disorders [15]. Peripheral blood smears, exhibiting variations in e.g., the size, color, shape of the red blood cells can provide diagnostic information on blood disorders including SCD[16]. In addition to diagnosis, inspection of blood smears is also frequently used for evaluation of treatment and routine monitoring of patients[17]. Preparation of these blood smear slides is rather straight-forward (i.e., can be performed by minimally-trained personnel), rapid and inexpensive. However, this method requires a trained expert to operate a laboratory microscope and perform manual analysis once the blood smear is prepared; the availability of such trained medical personnel for microscopic inspection of blood smears is limited in resource scarce settings, where the majority of people with SCD live[18]. In an effort to provide a solution to this bottleneck, deep learning-based methods have been previously used to classify[19] and segment[20] different types of red blood cells from digital images that were acquired using laboratory-grade benchtop microscopes equipped with oil-immersion objective lenses. However, these earlier works focused upon cell level detection, rather than slide level classification and therefore did not demonstrate patient level diagnosis or screening of SCD.

As an alternative to benchtop microscopes, smartphone-based microscopy provides a cost-effective and POC-friendly platform for microscopic inspection of samples, making it especially suitable for use in resource limited settings[21–23]. Smartphone microscopy has been demonstrated for a wide range of applications, including e.g., the imaging of blood cells[24,25], detection of viruses and DNA[26,27], quantification of immunoassays[28–31] and microplates[32] among many others [33–36]. Recently, machine learning approaches have also been applied to smartphone microscopy images for automated classification of parasites in soil and water[37,38].

Here we present a smartphone-based microscope and machine learning algorithms that together form a cost-effective, portable and rapid sickle cell screening framework, facilitating early diagnosis of SCD even in resource-limited settings. The mobile microscope (Figure 1) utilizes an opto-mechanical attachment, coupled to the rear camera of a smartphone, transforming it into a portable microscope using external parts that cost ~$60 in total. This compact microscope design has sub-micron spatial resolution[39] and weighs only 350 g including the smartphone itself.

Using this cost-effective mobile microscope, we performed slide-level automated diagnosis of SCD by rapidly classifying thousands of red blood cells within a large field-of-view using a deep learning-based framework that takes <7 sec to process a blood smear slide per patient. We blindly tested this approach using 96 blood smears (32 of which came from individual patients with SCD) and achieved ~98% accuracy together with an area-under the-curve (AUC) of 0.998. We believe that this platform provides a robust solution for cost-effective and rapid screening of SCD, making it especially promising for POC use in resource-limited settings.

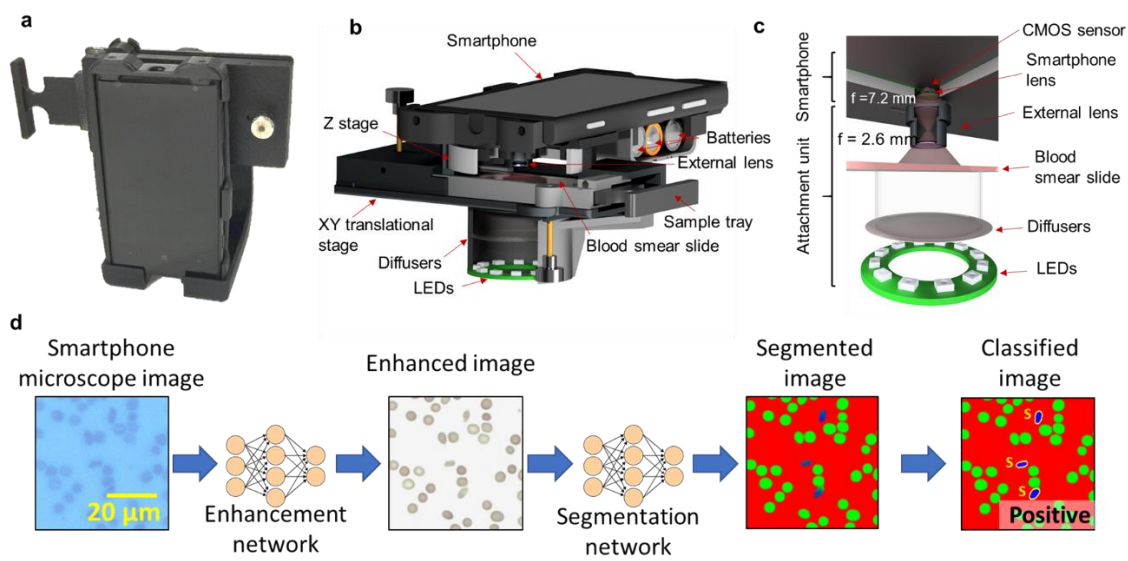

*Figure 1.* *a) A photograph of the smartphone-based brightfield microscope. Schematic illustration of b) the design of the portable microscope in detail and c) the light path. d) Deep learning workflow for sickle cell analysis.*

**Results**

The process of screening peripheral blood smears using our framework is illustrated in Figure 1 (d). Following the imaging of the patient slide with the smartphone-based microscope (Figure 1(a,b, and c)), these smartphone images were rapidly enhanced using a deep neural network as shown on the left part of Figure 1(d). This network was trained to transform the lower resolution, spatially and spectrally aberrated images of the smartphone microscope into enhanced images that are virtually equivalent to the images of the same samples captured using a higher numerical aperture (NA) laboratory-grade benchtop microscope. During the training phase (Methods section), which is a one-time effort, pairs of mobile microscopy images (input to the network) were cross-registered to the corresponding images of the same training samples, captured using a ×20/0.75 NA objective-lens with a ×2 magnification adapter (i.e., ×40 overall magnification), which were used as ground truth image labels.

This intermediate image transformation is essential for not only the automated blood smear analysis using a subsequent classification neural network (Figure 1(d)), but is also important for the improvement of smartphone microscopy images to a level that can be used by expert diagnosticians for manual inspection of the blood smears. Due to aberrations and lower resolution, the raw smartphone microscope images of red blood cells might be relatively difficult to interpret by human observers, who are traditionally trained on high-end benchtop microscopes. While our framework automatically performs cell classification and slide-level SCD diagnosis, any manual follow-up by a trained expert requires digital images that can be accurately interpreted. This is an important need satisfied by our image enhancement neural network that is specifically trained on blood smear samples to enhance the smartphone microscope images.

Another major advantage of this approach is that the enhancement network *standardizes* the microscopic images of blood smears, making it easier for the second classification network perform its task and more accurately classify the sickle cells. Therefore, the enhancement network improves both the *quality* and the *consistency* of the subsequent sickle cell classification network by performing an image standardization at its output. It also helps us to account for variations between images over time. In fact, our sample collection was performed over the course of 3 years, and the blood smear images were captured with different smartphones (using the same opto-mechanical attachment, same phone manufacturer and model number), resulting in some variations between the acquired mobile-phone images over time. The image enhancement network was trained with blood smear samples imaged over this time in order to account for these variations in the raw image quality of smartphone microscopy, standardizing the resolution and spatial as well as spectral features of the output images of the network.

Using the structural similarity index (SSIM)[40], we quantified the effectiveness of this image enhancement network on unique fields-of-view from the same slides that the network was trained with. The neural network improved the SSIM of our smartphone microscopy images from 0.601±0.017 (input) to 0.965±0.012 (output), where the mean and standard deviation were calculated for 8 smartphone microscope images, compared with the ground truth image labels acquired with a high NA benchtop microscope (see the Methods Section). This shows that after the neural network-based transformation, the intensity component of the smartphone images become highly similar to the benchtop microscope images. Examples of these image transformations can be seen in Figure 2, as well as in Figure S1(a), where direct comparisons between the network output and the ground truth benchtop microscope (0.75 NA) images are shown.

Following the image enhancement and standardization network output, a second classification network was then used to segment the enhanced images into three classes of objects: normal red blood cells, sickle cells, and background. Using the output of this network, each patient blood smear slide imaged by smartphone microscopy was automatically analyzed and screened for SCD using 5 different fields-of-view, each covering 0.51 mm × 0.51 mm, i.e., a total of ~1.25 mm$^2$ area of each blood smear was processed by the classification network, screening on average 9630 red blood cells per patient sample. Following this segmentation, the number of sickle cells and normal cells contained within each image were automatically counted. The patient slides were classified as being SCD positive if the average number of sickle cells within 5 images covering a 1.25 mm$^2$ field-of-view was above 0.5% of the total red blood cell count for that sample. This threshold was chosen based on the performance of the classification neural network in the validation dataset (separate from our blind testing data) to mostly account for sickle shaped healthy cells found in normal blood smears.

After the training phase, we confirmed the accuracy of our framework by blindly testing 96 blood smear slides that were never seen by our networks before. Covering 96 individual patients, 32 of these blood smears are from SCD patients and 64 of them are from healthy individuals. These slides were imaged using our smartphone microscope between 2016 and 2019, and were anonymously obtained from existing specimen at the UCLA Medical Center; the clinical diagnosis of each patient sample was used as the ground truth label for each slide. In this blinded test, our framework achieved ~98% accuracy across these 96 blood smears, where there was one false positive slide and one false negative slide. As for our misclassifications, one healthy blood smear was found to have a significantly higher average number of sickle cells (0.64%) than the remaining healthy blood smears; the one false negative sample was only slightly below our 0.5% threshold, while exhibiting a higher percentage of sickle cells than any of the remaining normal blood smears. The percentage of sickle cells measured by our platform for each one of these blood smear slides is also listed in Table S1, and several examples of patches from these test images, as they digitally pass through the networks, are also shown in Figure 2.

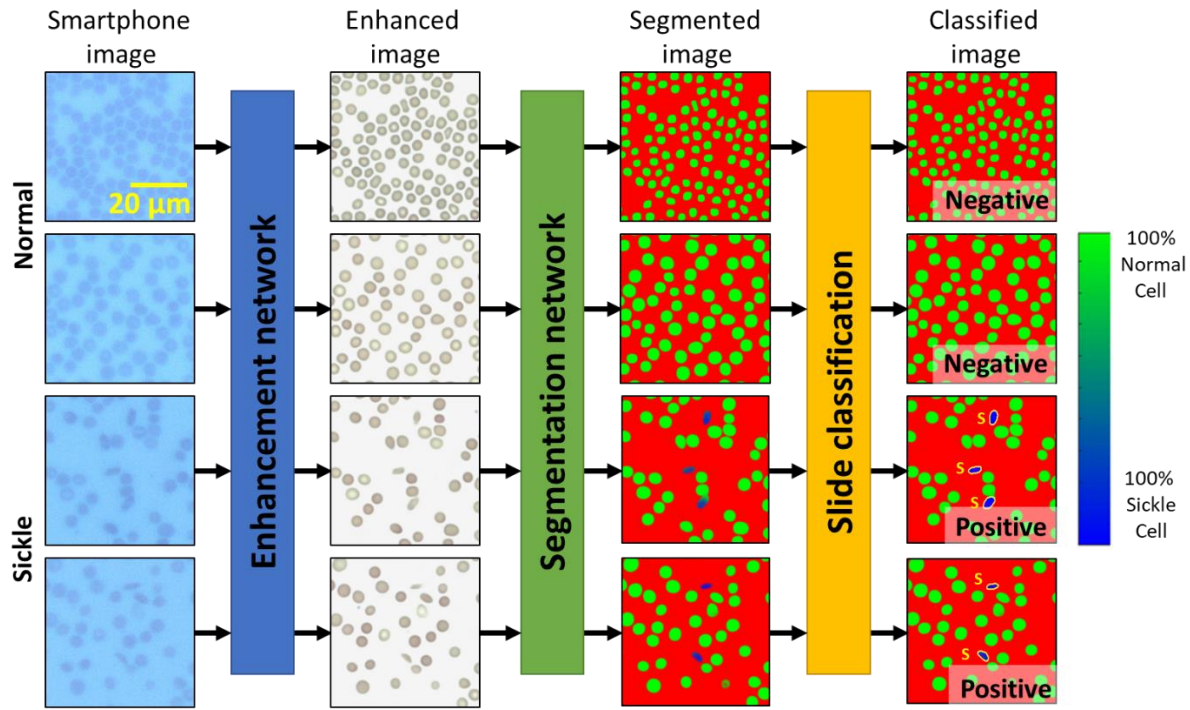

*Figure 2. Examples of patches from our test images that passed through the various steps of the automated sickle cell analysis framework. The smartphone microscopy images are first passed through an image enhancement and standardization network. Following this step, the images are segmented using a second, separate neural network. This segmentation network is in turn used to determine the number of normal and sickle cells within each image; 5 fields-of-view together covering ~1.25 mm$^2$ is automatically screened, having on average 9630 red blood cells to make a diagnosis for each patient blood smear.*

We also report the receiver operating characteristic (ROC) curve of our framework in Figure 3, which demonstrates how the SCD diagnosis accuracy can change depending on the threshold used to label the blood smear slide, achieving an AUC of 0.998.

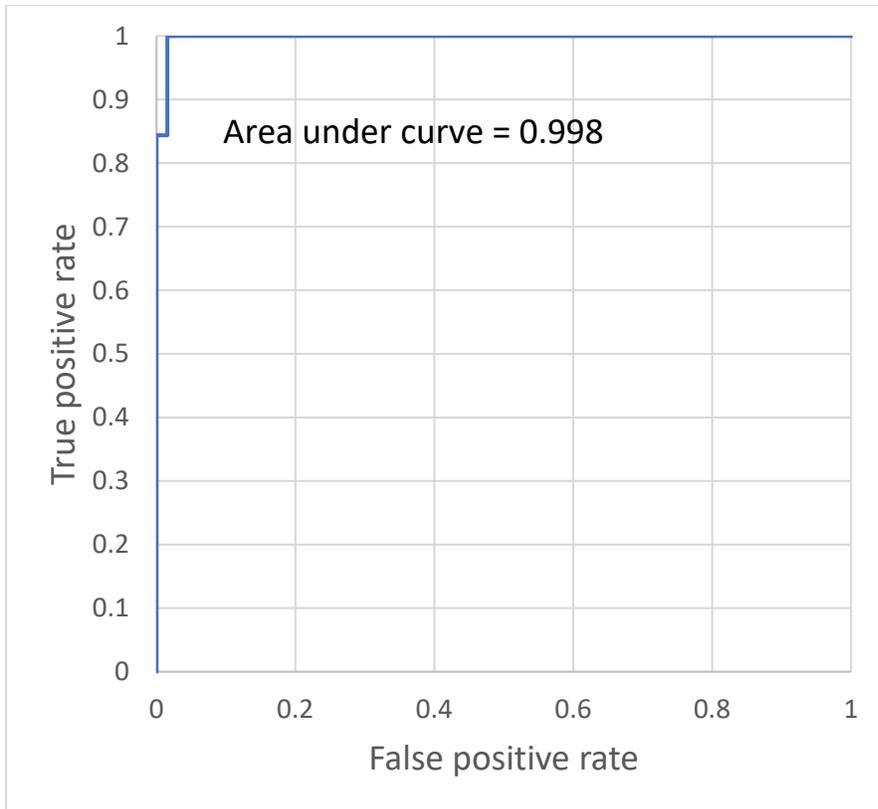

*Figure 3.* ROC curve demonstrating the false positive rate versus the true positive rate for our sickle cell detection framework.

**Discussion**

Through blind testing spanning 96 unique patient samples we have shown that the presented framework can consistently achieve high accuracy even using a limited training dataset. Similar to human diagnosticians examining blood smears under a microscope, screening through a large number of cells (on average 9630 red blood cells per patient sample) helped us achieve a high accuracy for automated diagnosis of SCD. In general, by using more training data containing a larger number of labeled sickle cell images, one can further improve our SCD detection accuracy and make it more efficient, requiring the capture of fewer smartphone images covering an even smaller field-of-view. On the other hand, acquiring a large training dataset with accurate labels at the single cell level is a challenge in itself; in fact, single cell level ground truth labels from blood smear samples clinically do not exist, and are not being recorded. Although one could perform single cell level molecular analysis on blood samples of patients, creating a sufficiently large scale image library backed up with such single cell level ground truth labels would be very costly and time consuming. As an alternative, one can use multiple diagnosticians to establish a clinical ground truth at the single cell level by statistically merging the decisions made by a panel of diagnosticians. A similar multi-diagnostician based decision approach has previously been used to detect malaria infected cells in blood smears, rather than relying on a single expert[41,42]. This approach mitigates the fact that single cell level labeling which is performed by a human can be highly subjective and inconsistent even for highly-trained experts[41]. Therefore, an accurate patient diagnosis can be difficult to perform with only a limited number of cells screened per slide, particularly with diseases such as SCD, where normal/healthy blood can also contain cells showing sickle cell like microscopic features. For example, children with normal blood were shown to have on average 0.28% sickle cells[43]. Due to this variation, a large number of cells must be screened per patient slide to better evaluate the rate of occurrence and make an accurate diagnosis.

Given these aforementioned challenges in obtaining large scale ground truth labeled image data from blood smears, our image enhancement and standardization network is particularly vital for SCD screening, helping the subsequent red blood cell segmentation network to better generalize sickle cell features and be efficient with limited training data by standardizing the input images that are fed into the classification network.

In this work, the blood smear slides were classified by measuring the percentage of sickle cells over a field-of-view of ~1.25 mm$^2$, covering on average 9630 red blood cells. Figure 4(a) reports how our diagnostic accuracy and AUC would change as a function of the number of cells that are screened per patient, further demonstrating the importance of inspecting a large sample field-of-view, and therefore a large number of cells for accurate SCD diagnosis (see the Methods section). Without inspecting an average of a few thousand red blood cells per patient slide, the accuracy of our automated SCD screening platform can drop significantly. Figure 4(b) also reports how the ROC curves are impacted as a function of the number of cells being screened per patient slide, confirming the same conclusion that both the sensitivity and the specificity of the test steadily drop as the number of inspected cells decreases.

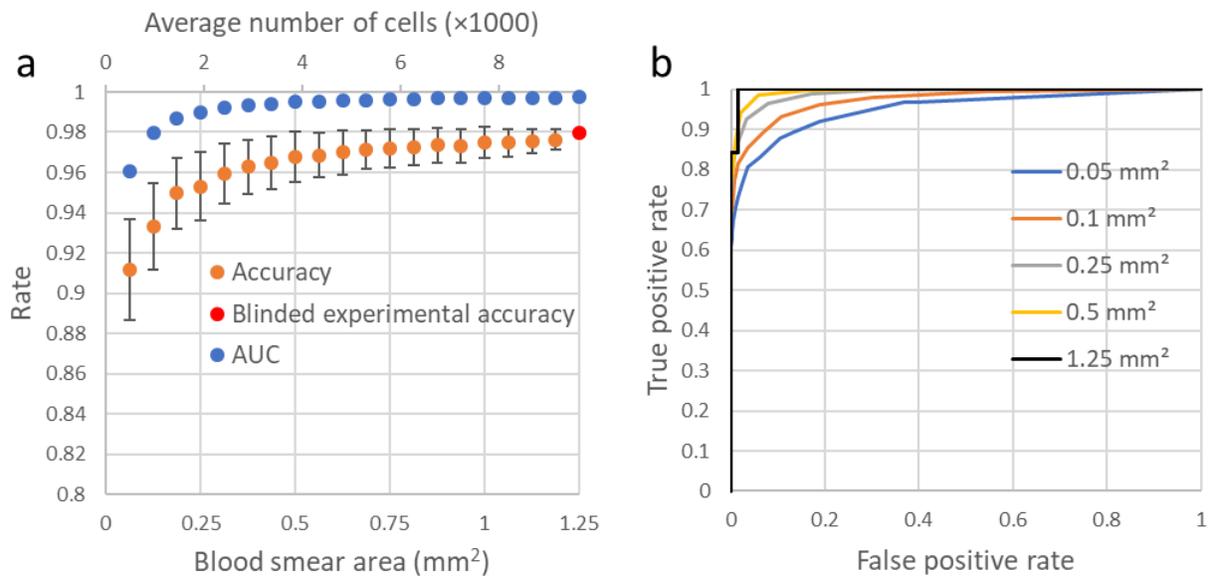

*Figure 4.* *a) Plot of how the accuracy and AUC change as a function the number of cells (and the blood smear area) inspected by our method. b) ROC curves for various simulated blood smear areas. These plots (except the 1.25 mm² one, which is our experimental result) are based on the average of 1000 Monte Carlo simulations performed by removing the red blood cells from the imaging fields-of-view at random to change the number of cells inspected by our method. As*

*the cells are relatively monodisperse, this random removal of the cells simulates a reduction in the inspected blood smear area per patient.*

Our results and analysis demonstrate that the presented method, enabled by smartphone microscopy and deep learning, is robust to perform SCD diagnosis by automated processing of blood smears. Furthermore, the test is rapid, cost-effective, and the required sample preparation is minimal, which is being routinely performed even in resource limited settings, resulting in hundreds of thousands of blood smears prepared per year just in sub-Saharan Africa[44]. Our method is also quite fast to compute an answer: each one of the 5 images passes through the neural networks in 1.37 seconds using a single Nvidia 1080 Ti GPU. This results in a total analysis time of 6.85 sec per patient, which is significantly faster than any manual inspection by experts. While in the current implementation the analysis is performed on a computer (which can be a local or remote server), a smartphone application could be also used to perform the processing on the phone itself with an increase in the slide processing time.

**Materials and Methods**

*Study design*

The objective of the research was to perform automated diagnosis of SCD using blood smear slides imaged with a smartphone-based microscope and analyzed by deep learning. The test dataset was made up of 96 unique patient samples involving 32 SCD thin blood smears and 64 normal thin blood smears. The blood smears were obtained from the UCLA Ronald Reagan Hospital, and no sample selection criterion was applied. Both the normal and SCD blood smears came from patients with a mix of gender and wide range of ages (<5 years to >60 years old).

*Ethics approval and consent to participate*

De-identified blood smears from existing human samples were obtained from the UCLA Ronald Reagan Hospital. No patient information, or any link to it, was disclosed to the research team. General consent for the samples to be used for research purposes was obtained. Due to these factors, no specific IRB from UCLA was required.

*Design of the smartphone-based brightfield microscope*

We used a Nokia Lumia 1020 smartphone attached to a custom-designed 3D-printed unit to capture images of the blood smear slides. An RGB light emitting diode (LED) ring (product no. 1643, Adafruit) was utilized to illuminate the sample in a transmission geometry and a microcontroller (product no. 1501, Adafruit) was used to adjust the illumination intensity. To ensure uniform illumination of the sample, a diffuser (product no. SG 3201, American Optic Supply, Golden, CO, USA) was placed in between the LEDs and the specimen. An external lens with a focal length of 2.6 mm (LS-40166 – M12xP0.5 Camera Lens) was used for magnification and was coupled to the rear camera of the smartphone. This design has a field of view of ~ 1 mm$^2$ per image. However, each one of our mobile phone images is cropped to the center ~0.5x0.5 mm$^2$ in order to avoid increased aberrations toward the edges of the field of view; per patient slide, we capture 5 independent images, covering a total of ~1.25 mm$^2$. The microscope is also equipped with a custom-designed manual translational stage to laterally move the sample. This stage, which was also 3D-printed, also contained a microscope slide holder. In total, the smartphone-based microscope weighs 350 g, including the phone itself, and the total cost of the device parts is ~ $60 (excluding the smartphone).

*Imaging of thin blood smears*

We used thin blood smear slides for image analysis. Our ground truth microscope images were obtained using a scanning benchtop microscope (model: Aperio Scanscope AT) at the Digital Imaging Laboratory of the UCLA Pathology Department. The standard smartphone camera application was used to capture the corresponding input images using the smartphone-based microscope, using auto focus, ISO 100, and auto exposure.

Areas of the samples captured using the smartphone microscope were co-registered to the corresponding fields-of-view captured using the benchtop microscope (please refer to "Image co-registration" in Methods section for details). Three board-certified medical doctors labelled the sickle cells within the images captured using the benchtop microscope using a custom-designed graphical user interface (GUI). As the images are co-registered, these labels were used to mark the locations of the sickle cells within the smartphone images, forming our training image dataset. We captured the images on the feathered edges of the blood smear slides, where the cells are dispersed as a monolayer.

Images from blood smears containing cells which have been scraped and damaged were excluded from the dataset, as the cut cells can appear similar to sickle cells (see e.g., Figure S2). One normal blood smear was accordingly excluded as we were unable to capture a sufficient number of usable fields-of-view due to the poor quality of the blood smear, with many scratches on its surface. Blood smears from four patients who were tested positive for SCD and were taking medicine for treatment were also excluded from the study since their smears did not contain sickle cells when viewed by a board-certified medical expert.

*Image co-registration*

The co-registration between the smartphone microscope images and those taken by the clinical benchtop microscope (NA=0.75) was done using a series of steps. For the first step, these images are scaled to match one another by bicubically down-sampling the benchtop microscope image to match the size of those taken by the smartphone. Following this, they are roughly matched using an algorithm which creates a correlation matrix between each smartphone image and the stitched whole slide image captured using the benchtop lab-grade microscope. The area with the highest correlation is the field of view which matches the smartphone microscope image and is cropped from the whole slide image. An affine transformation was then calculated using MATLAB's (Release R2018a, The MathWorks, Inc.) multimodal registration framework which extracts feature vectors and matches them to further correct the size, shift, shear, and account for any rotational differences[45]. Finally, the images were matched to each other using an elastic pyramidal registration algorithm to match the local features[39]. This step accounts for the spherical aberrations, which are extensive due to the nature of the inexpensive optics coupled to the smartphone camera. This algorithm co-registers the images at a subpixel level by progressively breaking the image up into smaller and smaller blocks and uses cross-correlation to align them.

*Image enhancement neural network*

Due to the variations among the images taken by the smartphone microscope, a neural network is used to standardize images and improve their quality in terms of spatial and spectral features. These variations stem from e.g., changing exposure time, aberrations (including defocus), chromatic aberrations due to source intensity instability, mechanical shifts, etc. Some examples of the image variations that these aberrations create can be seen in Figure S3. The quality of the images taken by a smartphone microscope can be improved and transformed so that they closely resemble those taken with a state-of-the-art benchtop microscope by using a convolutional neural network[39]. Our image normalization and enhancement network uses the U-net architecture as shown in Figure 5 (a)[46]. The U-net is made up of three "down-blocks" followed by three "up-blocks". Each one of these blocks is made up of three convolutional layers, which use a 3×3 convolution kernel and a stride of one. In the case of the down-block, the second of these layers increases the number of channels by a factor of two, while the second convolutional layer in the up-block reduces the number of channels by a factor of one quarter. The down-blocks are used to reduce the size of the images using an average pooling layer with a stride of two, so the network can extract and use features at different scales. The up-blocks return the images to the same size by bilinearly up-sampling the images by a factor of two. Between each of the blocks of the same size, skip connections are added to allow information to pass by the lower blocks of the U-net. Between the bottom blocks, a convolution layer is also added to allow processing of those large-scale features. The first convolutional layer of the network initially increases the number of channels to 32, while the last one reduces the number back to the 3 channels of the RGB color space to match the benchtop microscope images (ground truth).

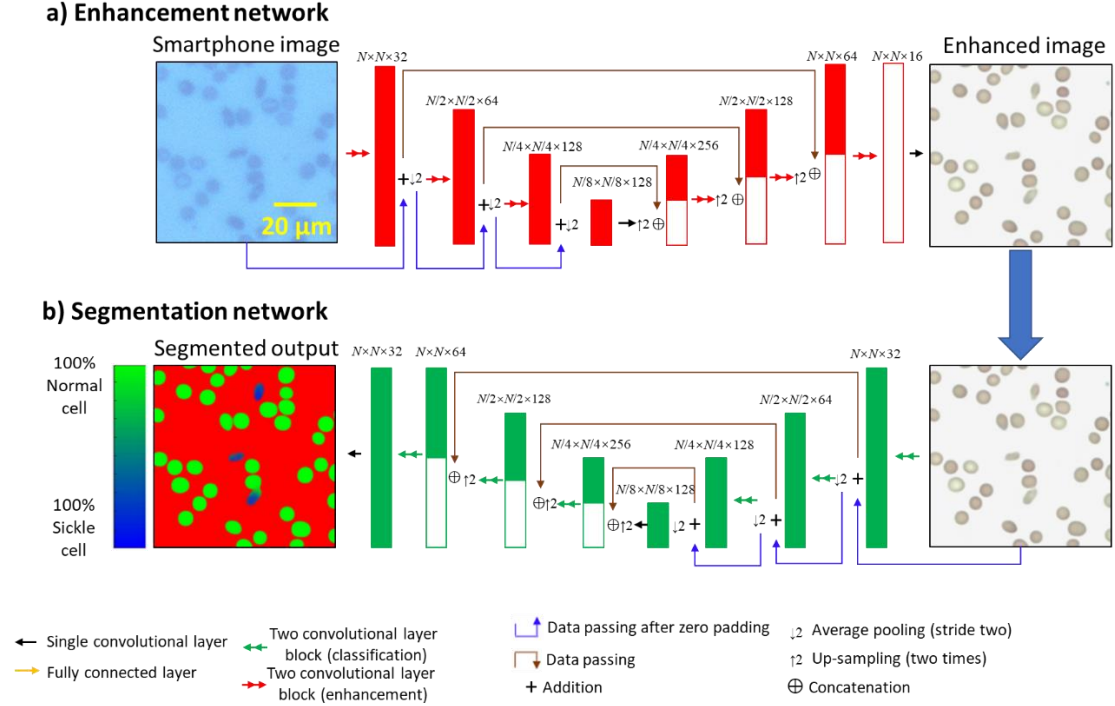

*Figure 5. Diagram detailing the network architecture for both a) the image enhancement network and b) the semantic segmentation network.*

The image enhancement network is trained using a combination of two loss functions, described by the equation:

$$L_{Network} = L_1\{z, G(x)\} + \lambda * TV(G(x)) \qquad (1)$$

where an $L_1$ (mean absolute error) loss function is used to train the network to perform an accurate transformation, while the total variation (TV) loss is used as a regularization term. $\lambda$ is a constant set to 0.03; this constant makes the total variation ~5% of the overall loss. *G(x)* represents image generated using the input image *x*. The $L_1$ loss can be described by the following equation[47]:

$$L_1\{z, G(x)\} = \frac{1}{N_{channels} \times M \times N} \sum_{n=1}^{N_{channels}} \sum_{i,j=1}^{M,N} |G(x)_{i,j,n} - z_{i,j,n}| \qquad (2)$$

Where $N_{channels}$ is the number of channels, $n$ is the channel number, $M$ and $N$ are the width and height of the image in pixels, and $i$, and $j$ are the pixel indices. The total variation loss is described by the following equation[48]:

$$TV(G(x)) = \frac{1}{N_{channels} \times M \times N} \sum_{n=1}^{N_{channels}} \sum_{i,j=1}^{M,N} \left( |G(x)_{i+1,j,n} - G(x)_{i,j,n}| + |G(x)_{i,j+1,n} - G(x)_{i,j,n}| \right) \quad (3)$$

The network was trained for 604000 iterations (118.5 epochs), with the data augmented through random flips and rotations of the training images by a multiple of 90 degrees.

For this image enhancement network training, there is *no need* for manual labeling of cells by a trained medical professional, and therefore this dataset can be made diverse very easily. Because of this, it can also be expanded upon quickly as all that is required is additional images of the slides to be captured by both microscopy modalities and co-registered with respect to each other. Therefore, the network was able to more easily cover the entire sample space to ensure accurate image normalization and enhancement. The training image dataset consisted of 520 image pairs coming from 10 unique blood smears. Each of these images have 1603×1603 pixels, and are randomly cropped into 128×128 pixel patches to train the network. Several examples of direct comparisons between the network's output and the corresponding field of view captured by the benchtop microscope can be seen in Figure S1(a).

*Mask creation for training the cell segmentation network*

Once the cells were labeled by board-certified medical experts and the images were co-registered, the cell labels were used to create a mask which constitutes the ground truth of the segmentation network; this mask creation process is a one-time training effort and used to train the cell segmentation neural network used in our work. These training masks were generated by thresholding the benchtop microscope images according to color and intensity to determine the locations of all the healthy and the sickle cells. The exact thresholds were chosen manually for each slide due to minor color variations between the blood smears; once again, this is only for the training phase. As the centers of some red blood cells were the same color as the background, holes in the mask were filled using MATLAB's imfill command, a morphological operator. Following this, the mask was eroded by four pixels in order to eliminate sharp edges and eliminate pixels misclassified due to noise. Any cell labeled by the medical expert as a sickle cell was set as a sickle cell while any unlabeled red blood cell was set as a normal cell for training purposes. White blood cells, platelets and the background were all labeled as a third background class. As the medical experts might have randomly missed some sickle cells within each field of view, a 128×128 region around each labeled sickle cell was cut out of the slide for training, reducing the unlabeled area contained within the training dataset. The remaining sections of the labeled slides were removed from the training dataset. At the end of this whole process, which is a one-time training effort, three classes are defined for the subsequent semantic segmentation training of the neural network: (1) sickle, (2) normal red blood cell, and (3) background.

*Semantic segmentation*

A second deep neural network is used to perform semantic segmentation of the blood cells imaged by our smartphone microscope. This network has the same architecture as the first image enhancement network (U-net). However, as this network performs segmentation, it uses the SoftMax cross entropy loss function to differentiate between the three classes (sickle cell, normal red blood cell, and background). In order to reduce the number of false positives as much as possible, the normal cell class is given twice the weight of the background and the sickle cells in the loss function. The overall loss function for the segmentation network, $L_{Segmentation}$, is described in equation 4:

$$L_{Segmentation} = -\frac{1}{M \times N} \sum_{i,j=1}^{M,N} a_{i,j,1} \log(p_{i,j,c=1}) + 2a_{i,j,2} \log(p_{i,j,c=2})$$

$$+ a_{i,j,3} \log(p_{i,j,c=3}) \quad (4)$$

where $M$ and $N$ are the number of pixels in an image, and $i$, and $j$ are the pixel indices as above. $a_{i,j,c}$ is the ground-truth binary label for each pixel (i.e., 1 if the pixel belongs to that class, 0 otherwise), and $c$ denotes the class number ($c=\{1,2,3\}$), where the first class represents the background, the second class is for healthy cells, and the third class is for sickle cells. The probability $p_{i,j,c}$ that a class $c$ is assigned to pixel $i,j$ is calculated using the softmax function:

$$p_{i,j,c} = \frac{\exp(y_{i,j,c})}{\sum_{k=1}^{3} \exp(y_{i,j,k})} \quad (5)$$

where $y$ is the output of the neural network.

A visual representation of the network architecture can be seen in Figure 5(b). Several examples of direct comparisons between the network's output at the single cell level and the corresponding field of view imaged by the clinical benchtop microscope can be seen in Figure S1(a).

The training dataset for this network was made up of 2660 sickle cell image patches (each 128×128 pixels) from a single blood smear slide, each one containing a unique labeled sickle cell. An additional 3177 image patches (each 128×128 pixels) coming from 15 unique slides containing solely normal cells were also used. Separate from our blind testing image dataset which involved 96 unique patients, 250 labeled 128×128-pixel sickle cell image patches and two 1500×1500-pixel images from healthy image slides were used as validation dataset for the network training phase. The classification algorithm was validated using these images alongside 5 unique fields-of-view from 10 additional blood smear slides of healthy patients.

*Classification of blood smear slides*

Once the images have been segmented by the second neural network, the number of total cells and sickle cells must be extracted. The algorithm first uses a threshold to determine which pixels are marked as cells. Areas where the *sum* of the sickle cell and normal cell probabilities is above 0.8 are considered to be part of a red blood cell, while areas below this threshold are considered as background regions. Connected areas which contain more than 100 pixels above the 0.8 threshold are then counted to determine the total number of cells. Sickle cells are counted using a similar methodology: connected areas where there are over 100 pixels above a sickle cell probability threshold of 0.15 were counted as sickle cells. This threshold is set to be low since significantly more number of healthy red blood cells is used to train the network. A slide is classified as being positive for sickle cell disease when the percentage of sickle cells among all the inspected cells (sickle and normal red blood cells) over a total field-of-view of ~1.25 $mm^2$ is above 0.5%. The 0.5% threshold was chosen using the validation image dataset, i.e., it was based on the network's performance in classifying the 10 healthy validation slides to account for false positives and the occurrence of sickle shaped cells in normal blood smears. Several examples of direct comparisons between the network's output and the ground truth labels for blindly tested regions of the labeled slides are shown in Figure S1(b).

*Structural similarity calculations*

The SSIM calculations were performed using only the brightness (Y) component of the YCbCr color space as we expect the intensity contrast component to remain similar, while the chroma components (Cb, Cr) to depend on other factors, including variability in the slide's staining. The calculations were performed upon 8 unique fields-of-view from the same slides which were used to train the enhancement network. SSIM is calculated using the equation:

$$SSIM(x,z) = \frac{(2\mu_x\mu_z + c_1)(2\sigma_{x,z} + c_2)}{(\mu_x^2 + \mu_z^2 + c_1)(\sigma_x^2 + \sigma_z^2 + c_2)} \qquad (6)$$

where *x* and *z* represent the two images being compared, as above. $\mu_x$ and $\mu_z$ represent the average values of *x* and *y* respectively, and $\sigma_x$ and $\sigma_z$ are the variance of *x* and *z*, and $\sigma_z$ is the covariance of *x* and *z*. $c_1$ and $c_2$ are dummy variables, which stabilize the division from a small denominator.

*Monte Carlo simulation details*

The Monte Carlo simulations reported in Figure 4 demonstrate how the accuracy of the presented technique changes as a function of the number of cells analyzed by our neural networks; these simulations were implemented by beginning with the full cell count from the 5 fields-of-view tested for each patient slide. This total cell count was reduced by randomly eliminating some of the cells to evaluate the impact of the number of cells analyzed on our accuracy. As the cells are relatively monodisperse, this random removal of red blood cells was used as an approximation of a reduction of the inspected blood smear area per patient. The results of 1000 simulations were averaged since the accuracy can fluctuate significantly, particularly at low numbers of cells. The total number of cells within the 5 fields-of-view that we used for SCD diagnosis varies from 4105 to 13989.

*Implementation details*

The neural networks were trained using Python 3.6.2 and the TensorFlow package version 1.8.0. The networks were trained and test images were processed on a desktop computer running Windows 10 using an Intel I9-7900X CPU, 64 GB of RAM and one of the computer's two GPUs (NVIDA GTX 1080 Ti). The enhancement network infers each field of view in 0.73 seconds while the classification network inference takes 0.64 seconds per field of view, taking a total of 6.85 sec to process the entire 1.25 $mm^2$ area of the blood smear. For both of the neural networks, the training image data were augmented by using random rotations and flipping.

**Acknowledgments:** The authors acknowledge the funding of Koc Group, NSF and HHMI. The authors also acknowledge Dr. Valerie Arboleda and Dr. Jonathan Armstrong of UCLA Health for their help labeling sickle cells.